\documentclass{PoS}

\usepackage{amsmath}
\usepackage{amsfonts}
\usepackage{braket}

\newcommand{\adjoint}[0]{\dagger}
\newcommand{\conjg}[0]{\ast}
\newcommand{\ee}[0]{\textrm{e}}
\newcommand{\trsp}[0]{\mathop{\textrm{tr}_\textrm{sp}}}

\title{Exploring the Roper wave function in Lattice QCD}

\ShortTitle{Exploring the Roper wave function in Lattice QCD}

\author{\speaker{Waseem Kamleh}%
\thanks{This research was undertaken with the assistance of resources at the
NCI National Facility in Canberra, Australia, provided through the
National Computational Merit Allocation Scheme and the University of Adelaide Partner Share.
This research is
supported by the Australian Research Council.} \\
       Special Research Centre for the Subatomic Structure of  Matter, \\
       University of Adelaide, Australia \\
       E-mail: \email{waseem.kamleh@adelaide.edu.au}}

\author{Derek B. Leinweber \\
       Special Research Centre for the Subatomic Structure of  Matter, \\
       University of Adelaide, Australia}

\author{Dale S. Roberts \\
       Special Research Centre for the Subatomic Structure of  Matter, \\
       University of Adelaide, Australia, and \\
       National Computational Infrastructure, \\
       Australian National University, Canberra, Australia}

\abstract{Using a correlation matrix analysis consisting of a variety
  of smearings, the CSSM Lattice collaboration has successfully
  isolated states associated with the Roper resonance and other
  "exotic" excited states such as the $\Lambda(1405)$ on the lattice
  at near-physical pion masses. We explore the nature of the Roper by
  examining the eigenvectors that arise from the variational analysis,
  demonstrating that the Roper state is dominated by the $\chi_1$
  nucleon interpolator and only poorly couples to $\chi_2.$ By
  examining the probability distribution of the Roper on the lattice,
  we find a structure consistent with a constituent quark model. In
  particular, the Roper $d$-quark wave function contains a single node
  consistent with a $2S$ state. A detailed comparison with constituent
  quark model wave functions is carried out, validating the approach
  of accessing these states by constructing a variational basis
  composed of different levels of fermion source and sink smearing.}

\FullConference{31st International Symposium on Lattice Field Theory - LATTICE 2013\\
		July 29 - August 3, 2013\\
		Mainz, Germany}

\begin{document}

\section{The Roper Resonance}

The first positive-parity excited state of the nucleon, known as the
Roper resonance, $N{\frac{1}{2}}^{+}$(1440 MeV) ${\rm P}_{11}$, has
presented a long-standing puzzle since its discovery in the 1960's due
to its lower mass compared to the adjacent negative parity,
$N{\frac{1}{2}}^{-}$(1535 MeV) ${\rm S}_{11}$, state.  In constituent
quark models with harmonic oscillator potentials, the lowest-lying
odd-parity state naturally occurs below the ${\rm P}_{11}$
state~\cite{Isgur:1977ef,Isgur:1978wd}. In nature the Roper resonance
is almost 100 MeV below the ${\rm S}_{11}$ state.

\begin{figure}
\centering
\includegraphics[height=0.6\textwidth,angle=90]{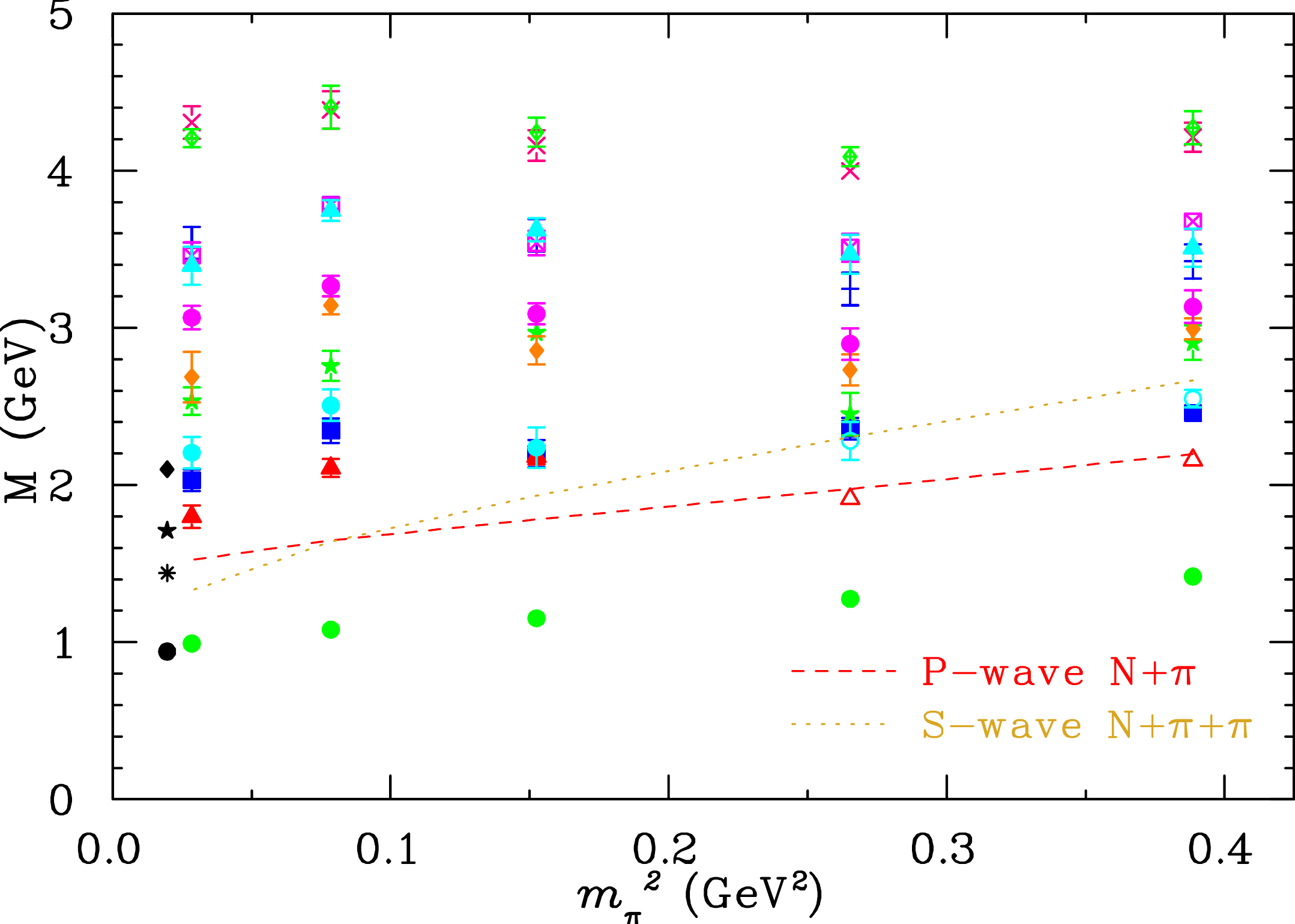}
\caption{\label{nucleon-spectrum}The even parity nucleon spectrum in
  full QCD from \cite{Mahbub:2010rm}, resulting from the superposition of two 8$\times$8
  correlation matrix analyses. The two lowest scattering channels are shown for reference.}
\end{figure}

Using variational techniques\cite{Michael198558, Lüscher1990222}, the
CSSM \cite{Mahbub:2009nr,Mahbub:2010jz,Mahbub:2010me} has successfully
isolated the elusive Roper \cite{Mahbub:2009aa,Mahbub:2010rm} and
$\Lambda(1405)$ \cite{Menadue:2011pd} resonances on the lattice.  The
highlights of the CSSM results for the even-parity nucleon spectrum in
Full QCD are shown in Figure~\ref{nucleon-spectrum}. These results
(and the full QCD results that follow below) were calculated on the
2+1 flavour non-perturbatively improved clover
configurations\cite{PhysRevD.79.034503} made available by the PACS-CS
collaboration via the ILDG\cite{Beckett:2009cb}. Critical to our
results is the construction of a large operator basis by considering
different amounts of gauge-invariant Gaussian
smearing\cite{Güsken1990361}.

We can see in Figure~\ref{nucleon-spectrum} that the first positive
parity excited state at the two heaviest pion masses has an energy
that is consistent with an $N+\pi$ multi-particle state, but as we
move to light quark masses the energy of this state shows significant
chiral curvature. We shall demonstrate that this state corresponds to
the Roper resonance. Specifically, we identify this state with the
$2S$ radial excitation of the nucleon. At the physical quark mass, the
lattice Roper state sits high when compared with the experimental
point at 1440 MeV. We argue that the finite volume of the lattice
causes the energy of the Roper to be pushed upwards away from the
physical value.

\section{Variational Analysis}

The success of the CSSM lattice program in identifying resonance
states on the lattice is founded upon variational techniques that
isolate individual excited states by constructing an operator basis
that couples to the states of interest and then considering their
cross-correlation matrix in order to diagonalise the operator
space. To access $N$ states of the spectrum, we require at least $N$
operators.

The parity-projected, two-point correlation function matrix for $\mathbf{p} = \mathbf{0}$ can be written as
\begin{equation}
G_{ij}^\pm(t) = \sum_{\mathbf{x}} \trsp(\Gamma_\pm \braket{\Omega | \chi_i(x) \smash{\overline\chi}_j(0) | \Omega}) = \sum_{\alpha = 0}^{N-1} \lambda_i^\alpha \smash{\overline\lambda}_j^\alpha \ee^{-m_\alpha t},
\end{equation}
where $\Gamma_\pm$ are the parity-projection operators and $\lambda_i^\alpha$ and $\smash{\overline\lambda}_j^\alpha$ are, respectively, the couplings of interpolators $\chi_i$ and $\smash{\overline\chi}_j$ at the sink and source to eigenstates $\alpha = 0, \ldots, N-1$ of mass $m_\alpha$. The idea now is to construct $N$ independent operators $\phi_i$ that isolate $N$ baryon states $\ket{B_\alpha}$; that is, to find operators $\smash{\overline\phi}^\alpha = \sum_{i=1}^N u_i^\alpha \smash{\overline\chi}_i$ and $\phi^\alpha = \sum_{i=1}^N v_i^{\alpha\conjg} \chi_i$ such that
\begin{align}
\braket{B_\beta, p, s | \smash{\overline\phi}^\alpha | \Omega} &= \delta_{\alpha\beta} \smash{\overline{z}}^\alpha \smash{\overline{u}}(\alpha, p, s),\textrm{ and} \nonumber \\
\braket{\Omega | \phi^\alpha | B_\beta, p, s} &= \delta_{\alpha\beta} z^\alpha u(\alpha, p, s),\label{phidef}
\end{align}
where $z^\alpha$ and $\smash{\overline{z}}^\alpha$ are the coupling strengths of $\phi^\alpha$ and $\smash{\overline\phi}^\alpha$ to the state $\ket{B_\alpha}$. It follows that 
\begin{equation}
G_{ij}^\pm(t) u_j^\alpha = \lambda_i^\alpha \smash{\overline{z}}^\alpha \ee^{-m_\alpha t},\label{Gu=lambdazexp}
\end{equation}
where, for notational convenience, we take the repeated Latin indices to be summed over while repeated Greek indices are not.

The only $t$ dependence in Eq.~\eqref{Gu=lambdazexp} is in the exponential term, so we immediately construct the recurrence relation $G_{ij}^\pm(t) u_j^\alpha = \ee^{-m_\alpha \Delta t} G_{ik}^\pm(t+\Delta t) u_k^\alpha$, which can be written as
\begin{equation}
(G^\pm(t+\Delta t))^{-1} G^\pm(t) \mathbf{u}^\alpha = \ee^{-m_\alpha \Delta t} \mathbf{u}^\alpha.
\end{equation}
This is an eigensystem equation for the matrix $(G^\pm(t+\Delta t))^{-1} G^\pm(t)$, with eigenvectors $\mathbf{u}^\alpha$ and eigenvalues $\ee^{-m_\alpha \Delta t}$.

Similarly, we can construct the associated left-eigensystem equation $\mathbf{v}^{\alpha\adjoint} G^\pm(t) (G^\pm(t+\Delta t))^{-1} = \ee^{-m_\alpha \Delta t} \mathbf{v}^{\alpha\adjoint}$, and then Eq.~\eqref{phidef} implies that
\begin{equation}
G^\pm_\alpha(t) := \mathbf{v}^{\alpha\adjoint} G^\pm(t) \mathbf{u}^\alpha = z^\alpha \smash{\overline{z}}^\alpha \ee^{-m_\alpha t}.
\end{equation}
Thus, the only state present in $G^\pm_\alpha(t)$ is $\ket{B_\alpha}$ of mass $m_\alpha$.

\section{Structure of the Roper}

\begin{figure}
\centering
\includegraphics[width=0.49\textwidth]{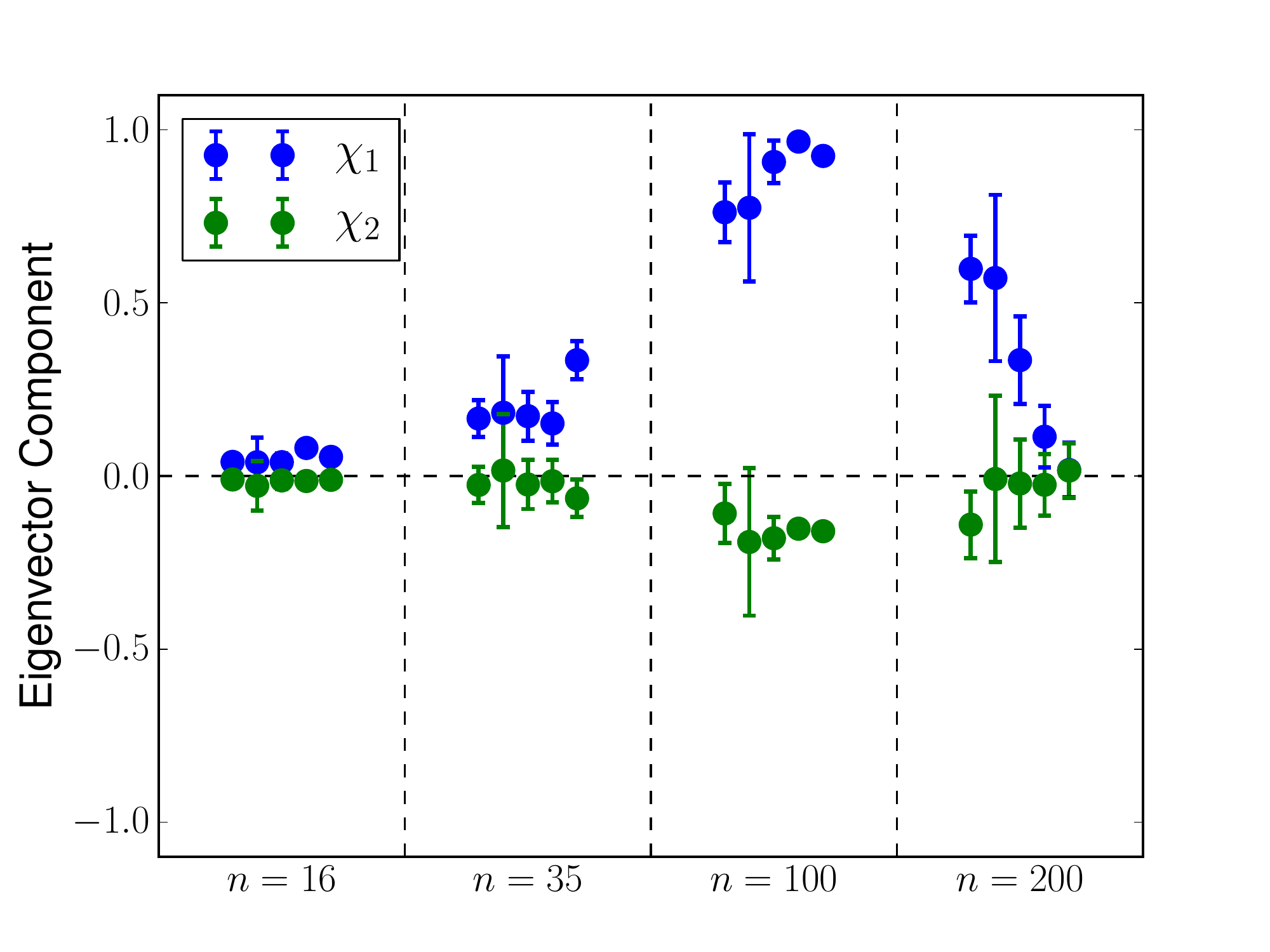}
\includegraphics[width=0.49\textwidth]{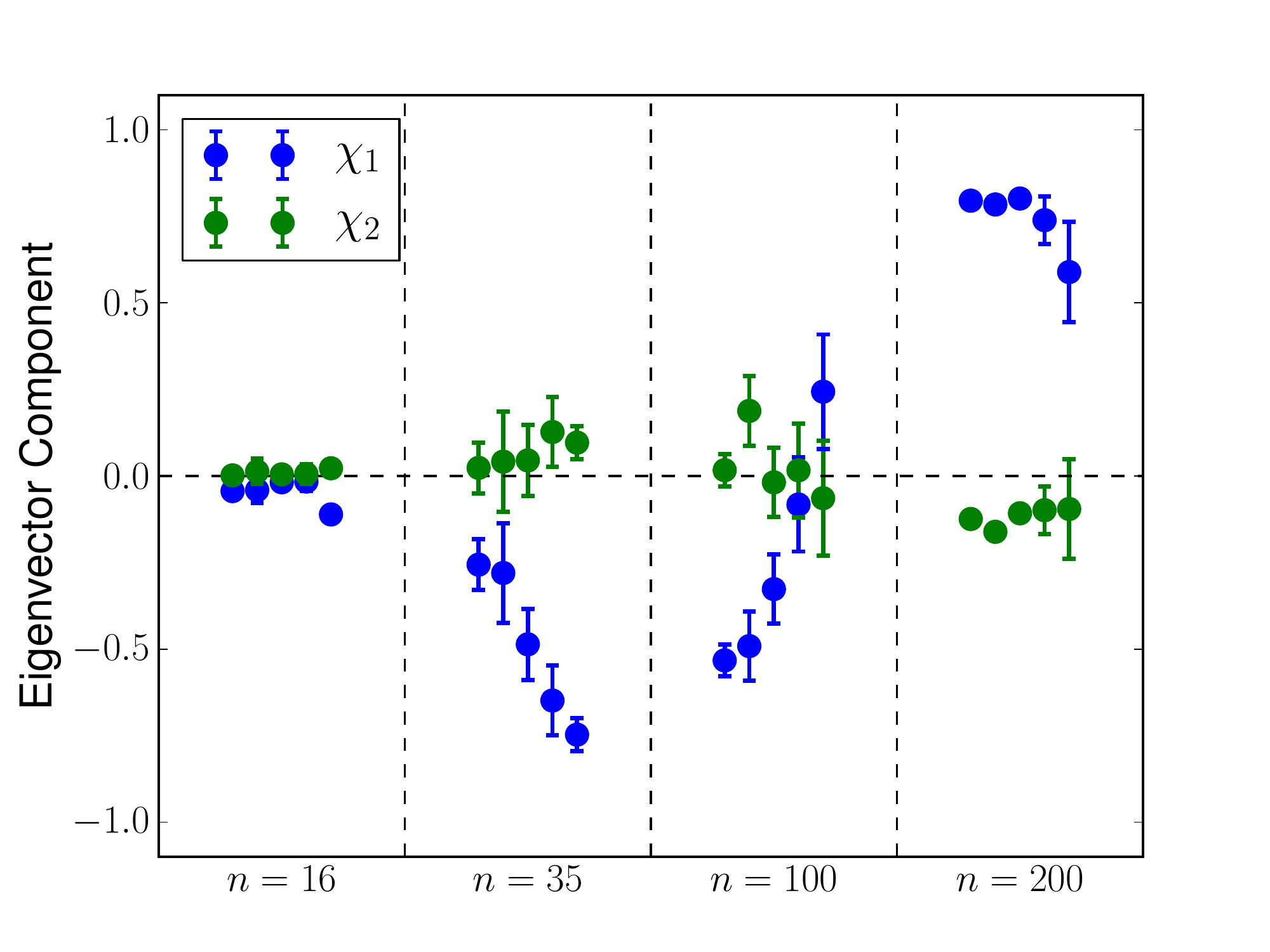}
\caption{\label{evecs}Eigenvector components for the nucleon ground state (left) and first excited state (right) resulting
  from an $8\times 8$ correlation matrix with $\chi_{1},\,\chi_{2}$
  interpolators. The blue points correspond to $\chi_1$ components and
  the green to $\chi_2.$ For each of the four smearing sweep values
  $n=16,35,100,200$ we display the eigenvector components at each of
  the five pion masses, with mass increasing from left to right.}
\end{figure}

Having constructed an appropriate operator basis, we can examine the
components of the right eigenvectors $\mathbf{u}^\alpha$ for a given
state $\alpha$ to see how each operator contributes to the formation
of that state. In Figure~\ref{evecs} we show the eigenvector
components for the ground and first excited states in the even parity
nucleon channel. We have four available smearings $n=16,35,100,200$
and two interpolating fields, $\chi_1$ (green points) and $\chi_2$
(blue points), giving us eight operators in total. For each operator,
the eigenvector components are shown in groups of five, corresponding
to each of the five pion masses $m_\pi=0.156, 0.293, 0.413, 0.572,
0.702$ GeV, with pion mass increasing from left to right.

Turning first to the left plot showing the ground state eigenmodes, we
see that the only significant contributions are from $\chi_1$ with
$100$ and $200$ smearing sweeps. Both these operators have positive
component, with the $n=200$ contribution increasing as the quark mass
decreases, which is consistent with the expectation that the width of
the nucleon ground state increases at lighter quark mass.

Now we examine the right plot showing the eigenmodes for the first
excited state.  Historically it was thought that the first
positive-parity excited state coupled strongly to $\chi_2.$ This
notion is rejected, as we see that the coupling of $\chi_2$ to the
first excited state is negligible at all four smearings. We observe
that the dominant contribution is $\chi_1$ at the largest smearing
$n=200$ with a positive sign, with opposite sign contributions from a
varying mixture of $n=35,100.$ 

The combination of Gaussians of different width but with opposite sign
suggests the possibility of a nodal structure for the first excited
state. In order to explore this notion further, we examine the
structure of the nucleon and its excitations on the lattice via the
Bethe-Salpeter amplitude, referred to hereafter simply as the wave function.
As detailed in \cite{Roberts2013164,Roberts:2013oea}, the baryon wave function is built by giving each quark field in the
annihilation operator a spatial dependence, 
\begin{equation}
\chi_1(\vec{x},\vec{y},\vec{z},\vec{w}) = \epsilon^{abc}\, (\,
  u_a^T(\vec{x}+\vec{y})\, C\gamma_5\, d_b(\vec{x}+\vec{z})\, )\,
  u_c(\vec{x}+\vec{w}),
\end{equation}
while the creation operator remains local. The resulting construction
is gauge-dependent, so we choose to fix to Landau gauge. Having
established that the contribution of $\chi_2$ is negligible to the
ground and first excited state, we obtain the right eigenvectors from
a $4\times 4$ correlation matrix analysis using $\chi_1$ only, at the
four different smearings. The non-local sink operator cannot be
smeared, so we then construct the baryonic wave functions using the
right eigenvector $u^\alpha$ only. The position of the $u$ quarks is
fixed at the origin and we measure the $d$ quark probability
distribution at the lightest quark mass $m_\pi = 156\text{ MeV}.$

\begin{figure}
\centering
\includegraphics[width=0.49\textwidth]{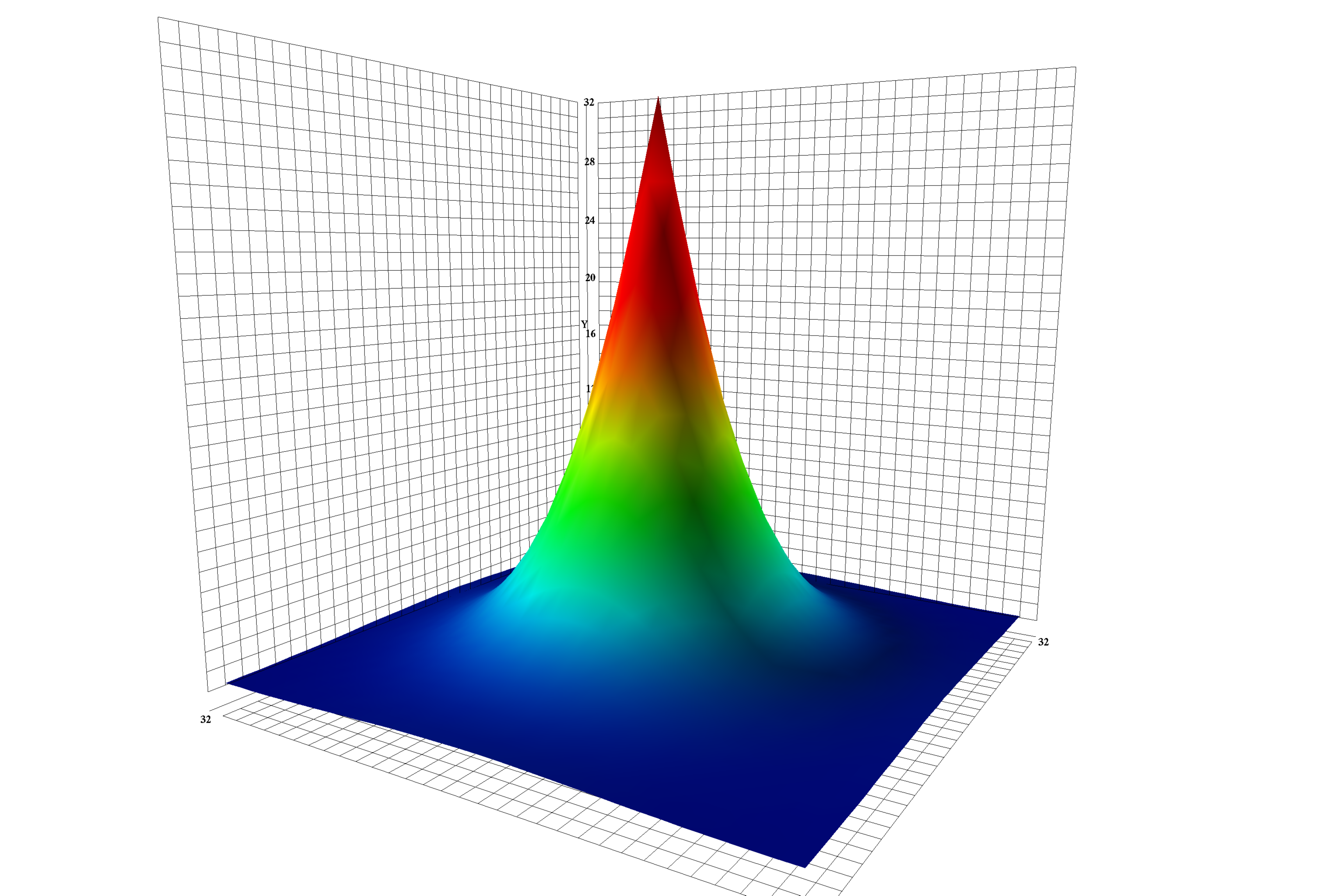}
\includegraphics[width=0.49\textwidth]{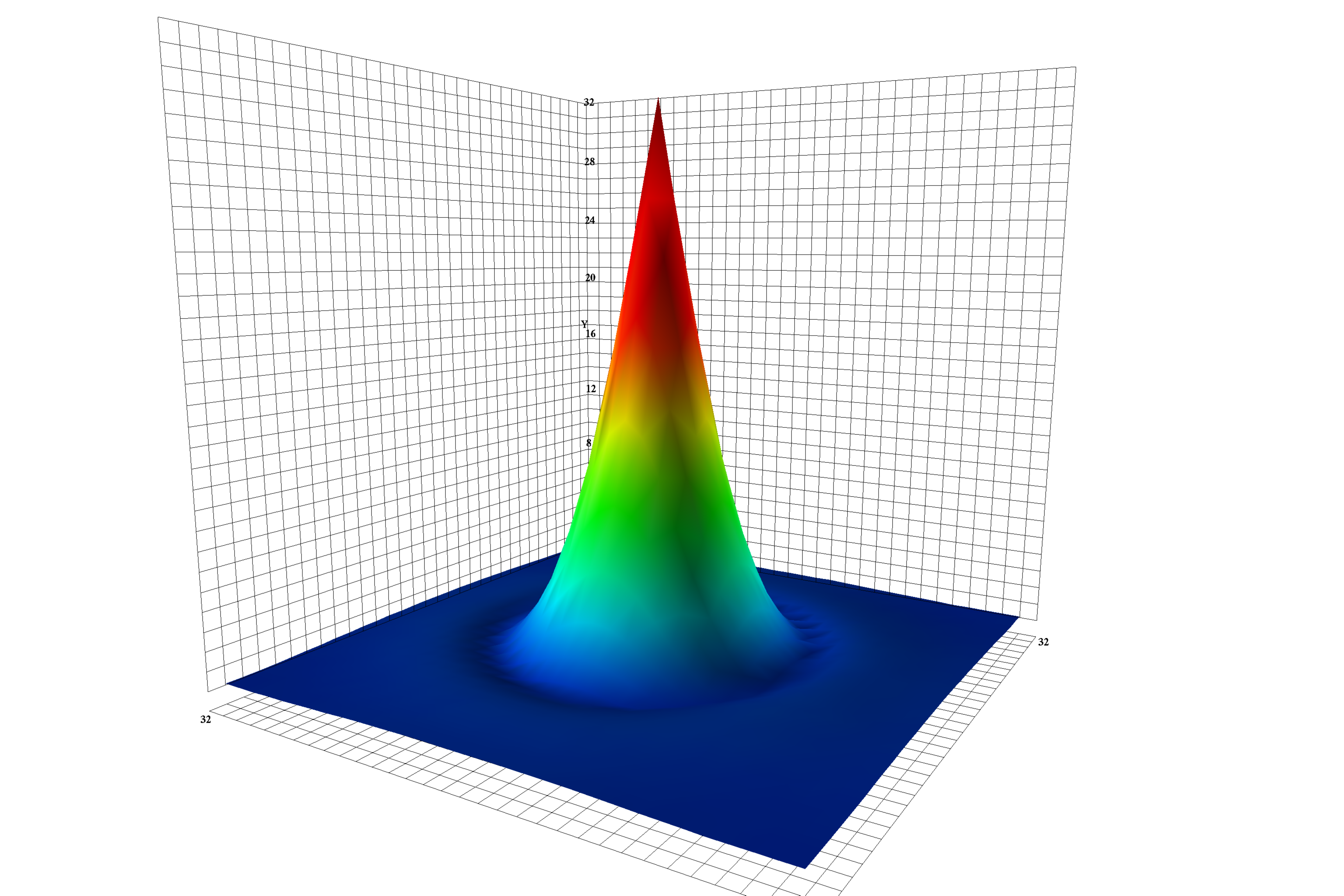}
\caption{\label{surface}The probability distribution of the $d$ quark
  about the two $u$ quarks at the origin in the ground state (left)
  and first excited state (right). The darkened ring around the peak
  in the first excited state indicates a node in the probability
  distribution, consistent with a $2S$ state.}
\end{figure}

The left plot of Figure~\ref{surface} shows the $d$-quark probability
distribution for the nucleon ground state. Recall from the eigenmode
analysis that this state is formed from $\chi_1,$ with the two largest
smearings both contributing with a positive sign. In this context, the
variational analysis is automatically constructing the nucleon ground
state by adjusting the mixture of the $n=100$ and $n=200$ operators in
order to create a Gaussian of the appropriate width. This
interpretation is verified by the simple structure of the ground state
wave function.

In contrast, the combination of a wide and narrow Gaussian with
opposite sign in the first excited state should yield a nodal
structure, and this is indeed what we observe in the right hand plot
of Figure~\ref{surface}. The probability distribution of the $d$-quark
in the first excited state clearly shows a single node, as one would
expect from a $2S$ radial excitation. 

\begin{figure}
\centering
\null\hfill\includegraphics[height=0.42\textwidth,angle=90]{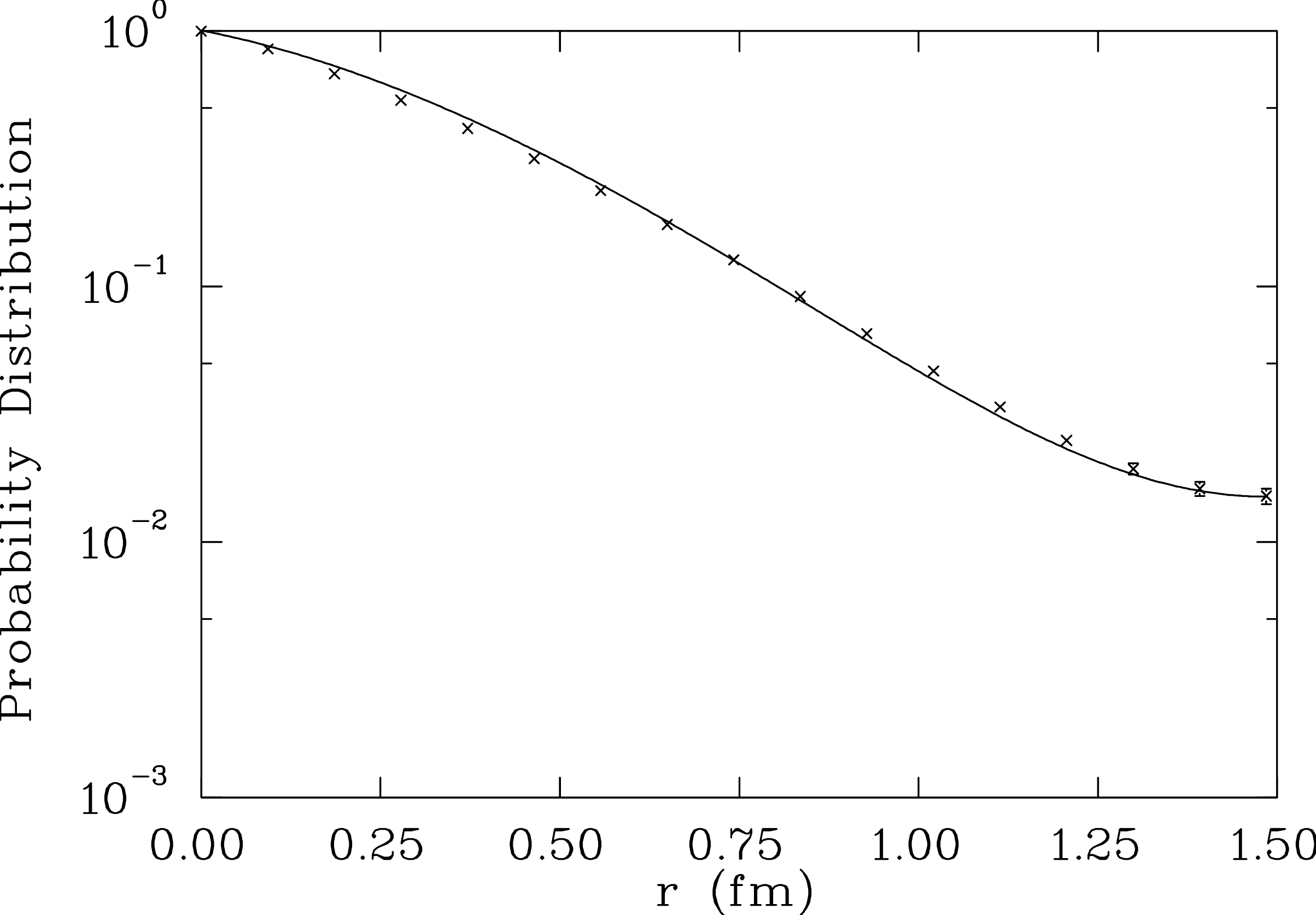}\hfill
\includegraphics[height=0.42\textwidth,angle=90]{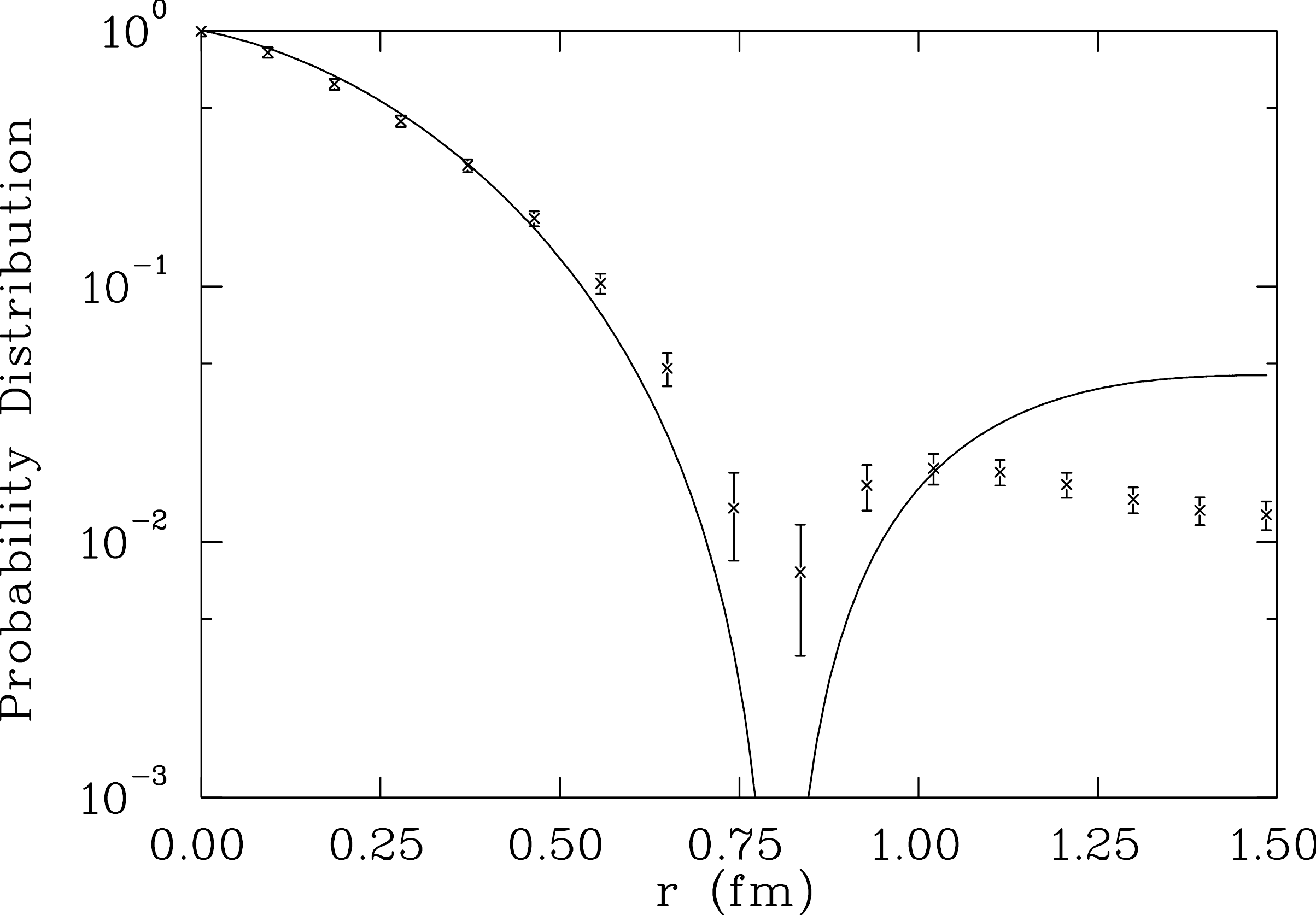}\hfill\null
\caption{\label{model}Comparison of the ground state (left) and first excited state (right) $d$-quark radial probability
  distribution from our lattice QCD calculation (crosses) with the
  quark model (solid curve). The quark model predicts the excited state node in
  approximately the correct location, but deviates at the boundary.}
\end{figure}

We can gain deeper insight into the first excited state structure by
comparing the radial wave function calculated on the lattice with a
non-relativistic constituent quark model, based on a
one-gluon-exchange motivated Coulomb + ramp potential \cite{Bhaduri:1980fd}. The
radial Schrodinger equation is solved with boundary conditions
relevant to the lattice data, that is, the derivative of the wave
function is set to vanish at a distance $L_x/2.$

In order to fix the model parameters, we use a two parameter fit to
the ground state lattice radial wave function.  This yields a string
tension value of $\sqrt{\sigma} = 400\text{ MeV},$ with a constituent
quark mass of $m_q = 360\text{ MeV}.$ These parameters are then held
fixed for the excited states, and we can compare the quark model
predictions with the lattice data.

In Figure~\ref{model}, we display the radial wave function predicted
by the quark model along with the lattice data. The left plot shows
that the quark model is able to fit the lattice ground state
relatively well. In the right plot we can now compare the quark model
prediction with first excited state radial wave function. We see that
the quark model successfully predicts the position of the node, but
disagrees with the lattice data as we approach the boundary.

\begin{figure}
\centering
\includegraphics[width=0.75\textwidth]{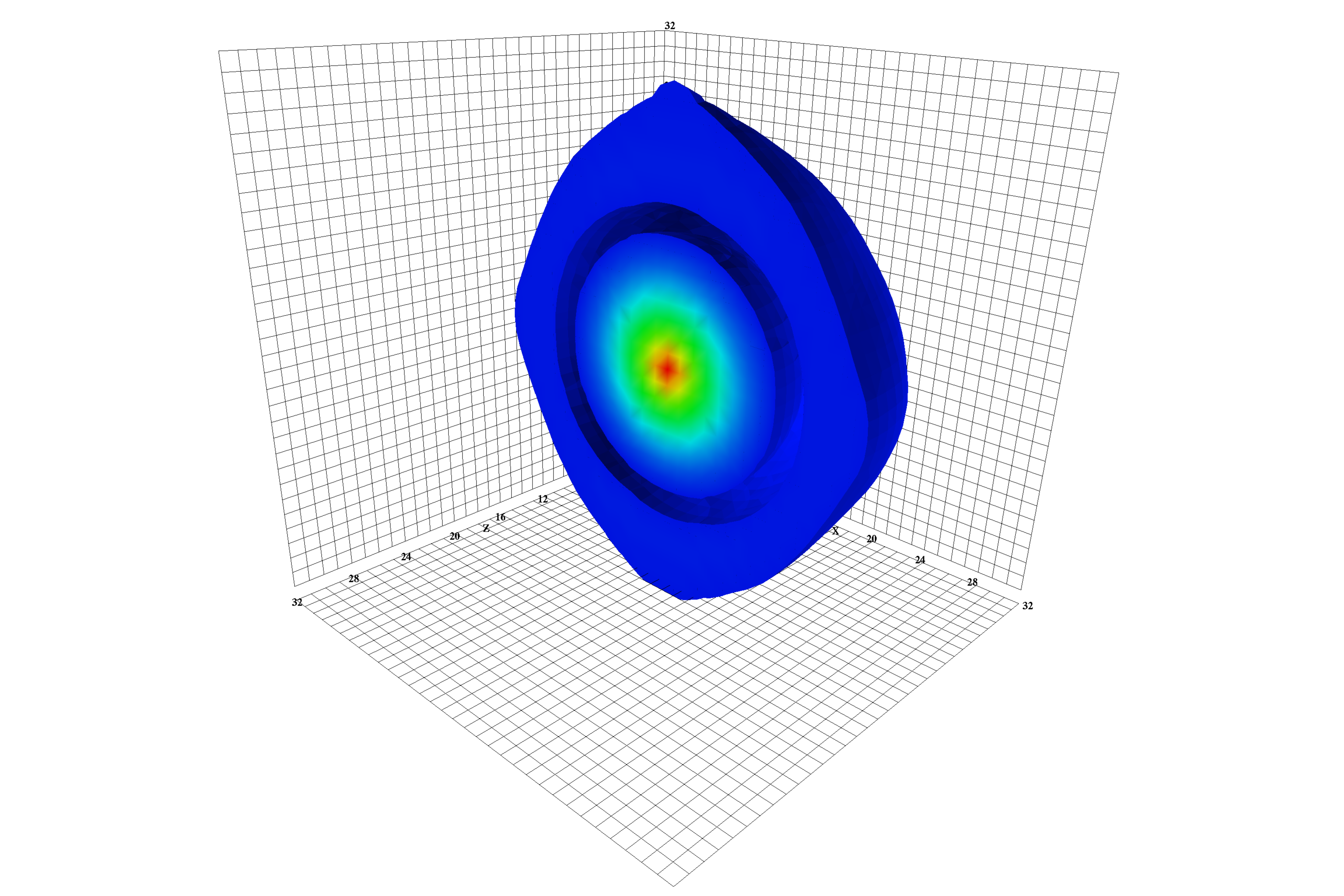}
\caption{\label{IsovolS2}The isovolume of the probability distribution of the $d$
  quark in the first excited state. The outer edge can be seen to be affected
  by the boundary, indicating a necessary finite-volume effect
  associated with multi-particle components of the state.} 

\end{figure}

Turning to Figure~\ref{IsovolS2}, we examine an isovolume display of
the $d$-quark probability distribution for the Roper. The node
structure is clearly visible, and within the node we can see that the
distribution has the spherical symmetry that one would expect from a
$2S$ radial excitation. However, the outer shell displays significant
deviation from spherical symmetry. The finite volume of the lattice
and periodic boundary conditions combine to distort what should be a
circular shell boundary into a ``rounded diamond'' shape. This
strongly suggests that finite volume effects are responsible for the
energy of the lattice Roper state sitting high compared to the
experimental value of 1440 MeV. These finite volume effects are also
the most likely cause of the significant disagreement with the quark
model radial wave function near the lattice boundary.

\section{Conclusions}

Using a variational analysis consisting of operators constructed with
a variety of Gaussian smearings, we find a low-lying positive parity
excited state that has a structure consistent with a $2S$ radial
excitation of the nucleon, which can therefore be identified as the
lattice realisation of the Roper resonance. The radial wave function
of the Roper is consistent with the predictions of a constituent quark
model. It is clear that at the lightest quark mass there are finite
volume effects present in the lattice results, which may cause the
state to sit high compared to the mass of the $N(1440)$ resonance in
Nature. Future avenues to explore include a study of the lattice
volume dependence of the Roper mass, examining the role that
multi-hadron operators can play in constructing a correlation matrix
analysis, and probing the electromagnetic structure of octet baryon 
excitations \cite{Boinepalli:2006xd}.



\providecommand{\href}[2]{#2}\begingroup\raggedright\endgroup


\end{document}